\documentclass[11pt]{article}
\usepackage{amssymb,amsfonts}
\usepackage{graphics,psboxit,amsmath}
\usepackage{subfigure}
\usepackage{graphicx}
\usepackage{verbatim}
\usepackage{color}
\usepackage{hyperref}
\hypersetup{colorlinks}

\definecolor{darkred}{rgb}{1,0,0}
\definecolor{darkgreen}{rgb}{0,0.5,0}
\definecolor{darkblue}{rgb}{0,0,1}
\definecolor{orange}{rgb}{1,0.5,0}
\definecolor{green}{rgb}{0,1,0}
\definecolor{purple}{rgb}{.5,0,1}

\hypersetup{ colorlinks,
linkcolor=darkred,
filecolor=darkgreen,
urlcolor=darkblue,
citecolor=darkblue,
linktocpage=true }

\definecolor{markcolor}{rgb}{.25,0,1}

\numberwithin{equation}{section}
\newtheorem{definition}{Definition}[section]

\newtheorem{proposition}[definition]{Proposition}

\newcommand{\ie}{{\it i.e.}\ }

\definecolor{markcolor2}{rgb}{1,0,0}

\definecolor{markcolor3}{rgb}{0,1,0}

\def\hybrid{\topmargin 0pt    \oddsidemargin 0.1in 
        \headheight 0pt \headsep 0pt
        \textwidth 16.0cm      
        \textheight 22,2cm       
        \marginparwidth .875in
        \parskip 5pt plus 1pt   \jot = 1.5ex}

\hybrid

\catcode`\@=11

\def\marginnote#1{}
\newcount\hour
\newcount\minute
\newtoks\amorpm
\hour=\time\divide\hour by60 \minute=\time{\multiply\hour by60
\global\advance\minute by-\hour}
\edef\standardtime{{\ifnum\hour<12 \global\amorpm={am}%
        \else\global\amorpm={pm}\advance\hour by-12 \fi
        \ifnum\hour=0 \hour=12 \fi
        \number\hour:\ifnum\minute<10 0\fi\number\minute\the\amorpm}}
\edef\militarytime{\number\hour:\ifnum\minute<10 0\fi\number\minute}

\def\draftlabel#1{{\@bsphack\if@filesw {\let\thepage\relax
   \xdef\@gtempa{\write\@auxout{\string
      \newlabel{#1}{{\@currentlabel}{\thepage}}}}}\@gtempa
   \if@nobreak \ifvmode\nobreak\fi\fi\fi\@esphack}
        \gdef\@eqnlabel{#1}}
\def\@eqnlabel{}
\def\@vacuum{}
\def\draftmarginnote#1{\marginpar{\raggedright\scriptsize\tt#1}}

\def\draft{\oddsidemargin -.5truein
        \def\@oddfoot{\sl preliminary draft \hfil
        \rm\thepage\hfil\sl\today\quad\militarytime}
        \let\@evenfoot\@oddfoot \overfullrule 3pt
        \let\label=\draftlabel
        \let\marginnote=\draftmarginnote
   \def\@eqnnum{(\theequation)\rlap{\kern\marginparsep\tt\@eqnlabel}%
\global\let\@eqnlabel\@vacuum}  }

\def\draft2{
        \def\@oddfoot{\sl preliminary draft \hfil
        \rm\thepage\hfil\sl\today\quad\militarytime}
        \let\@evenfoot\@oddfoot \overfullrule 3pt
        \let\label=\draftlabel
        \let\marginnote=\draftmarginnote
   \def\@eqnnum{(\theequation)\rlap{\kern\marginparsep\tt\@eqnlabel}%
\global\let\@eqnlabel\@vacuum}  }

\def\preprint{\twocolumn\sloppy\flushbottom\parindent 2em
        \leftmargini 2em\leftmarginv .5em\leftmarginvi .5em
        \oddsidemargin -.5in    \evensidemargin -.5in
        \columnsep .4in \footheight 0pt
        \textwidth 10.in        \topmargin  -.4in
        \headheight 12pt \topskip .4in
        \textheight 6.9in \footskip 0pt
        \def\@oddhead{\thepage\hfil\addtocounter{page}{1}\thepage}
        \let\@evenhead\@oddhead \def\@oddfoot{} \def\@evenfoot{} }

\def\numberbysection{\@addtoreset{equation}{section}
        \def\theequation{\thesection.\arabic{equation}}}

\def\underline#1{\relax\ifmmode\@@underline#1\else
        $\@@underline{\hbox{#1}}$\relax\fi}

\def\titlepage{\@restonecolfalse\if@twocolumn\@restonecoltrue\onecolumn
     \else \newpage \fi \thispagestyle{empty}\c@page\z@
        \def\thefootnote{\fnsymbol{footnote}} }

\def\endtitlepage{\if@restonecol\twocolumn \else \newpage \fi
        \def\thefootnote{\arabic{footnote}}
        \setcounter{footnote}{0}}  

\catcode`@=12 \relax

\def\figcap{\section*{Figure Captions\markboth
        {FIGURECAPTIONS}{FIGURECAPTIONS}}\list
        {Figure \arabic{enumi}:\hfill}{\settowidth\labelwidth{Figure
999:}
        \leftmargin\labelwidth
        \advance\leftmargin\labelsep\usecounter{enumi}}}
 \relax
\def\tablecap{\section*{Table Captions\markboth
        {TABLECAPTIONS}{TABLECAPTIONS}}\list
        {Table \arabic{enumi}:\hfill}{\settowidth\labelwidth{Table
999:}
        \leftmargin\labelwidth
        \advance\leftmargin\labelsep\usecounter{enumi}}}
 \relax
\def\reflist{\section*{References\markboth
        {REFLIST}{REFLIST}}\list
        {[\arabic{enumi}]\hfill}{\settowidth\labelwidth{[999]}
        \leftmargin\labelwidth
        \advance\leftmargin\labelsep\usecounter{enumi}}}
 \relax
\makeatletter
\newcounter{pubctr}
\def\publist{\@ifnextchar[{\@publist}{\@@publist}}
\def\@publist[#1]{\list
        {[\arabic{pubctr}]\hfill}{\settowidth\labelwidth{[999]}
        \leftmargin\labelwidth
        \advance\leftmargin\labelsep
        \@nmbrlisttrue\def\@listctr{pubctr}
        \setcounter{pubctr}{#1}\addtocounter{pubctr}{-1}}}
\def\@@publist{\list
        {[\arabic{pubctr}]\hfill}{\settowidth\labelwidth{[999]}
        \leftmargin\labelwidth
        \advance\leftmargin\labelsep
        \@nmbrlisttrue\def\@listctr{pubctr}}}
 \relax
\makeatother

\def\be{\begin{equation}}
\def\ee{\end{equation}}
\def\ba{\begin{eqnarray}}
\def\ea{\end{eqnarray}}

\def\del{\partial}

\def\mU{\mathbb U}
\def\mV{\mathbb V}
\def\mX{\mathbb X}
\def\mY{\mathbb Y}
\def\mT{\mathbb T}

\newcommand{\RR}{\mbox{${\mathbb R}$}}
\newcommand{\NN}{\mbox{${\mathbb N}$}}
\newcommand{\CC}{\mbox{${\mathbb C}$}}
\newcommand{\1}{\mbox{\hspace{.0em}1\hspace{-.24em}I}}

\def\bt{\mathbf t}

\def\cH{{\cal H}}
\def\cL{{\cal L}}

\def\cP{{\cal P}}

\def\IR{\relax{\rm I\kern-.18em R}}

\def\bse{\begin{small}\begin{equation*}}
\def\ese{\end{equation*}\end{small}}

\begin{document}


\renewcommand{\theequation}{\thesection.\arabic{equation}}
\csname @addtoreset\endcsname{equation}{section}

\newcommand{\eqn}[1]{(\ref{#1})}

\begin{titlepage}
\begin{center}
\strut\hfill

{\Large \bf Lagrangian and Hamiltonian structures in an integrable hierarchy and space-time duality}

\vskip 0.5in

{\bf Jean Avan$^{a}$, Vincent Caudrelier$^{b}$, Anastasia Doikou$^{c}$ and Anjan Kundu$^{d}$}
\vskip 0.05in
\vskip 0.02in
\noindent
$^{a}${\footnotesize Laboratoire de Physique Th\'eorique et Mod\'elisation (CNRS UMR 8089),\\
Universit\'e de Cergy-Pontoise, F-95302 Cergy-Pontoise, France}
\\[3mm]
\noindent
$^{b}${\footnotesize Department of Mathematics, City University London,\\ Northampton Square, EC1V 0HB London, United Kingdom}
\\[3mm]
\noindent
$^{c}${\footnotesize Department of Mathematics, Heriot-Watt University,\\
EH14 4AS, Edinburgh, United Kingdom}
\\[3mm]
\noindent
$^{d}${\footnotesize Saha Institute of Nuclear Physics, Theory Division, Kolkata, India}
\\[3mm]
\noindent
\vskip .1cm

\end{center}

\vspace{1cm}

\centerline{\bf Abstract}

We define and illustrate the novel notion of dual integrable hierarchies, on the example of the nonlinear Schr\"odinger (NLS) hierarchy. 
For each integrable nonlinear evolution equation (NLEE) in the hierarchy, dual integrable structures are characterized by 
the fact that the zero-curvature representation of the NLEE can be realized 
by two Hamiltonian formulations stemming from two distinct choices of the configuration space, yielding two inequivalent Poisson structures on the 
corresponding phase space and two distinct 
Hamiltonians. This is fundamentally different from the standard bi-Hamiltonian or generally multitime structure.
The first formulation chooses purely space-dependent fields as configuration space; it yields the 
standard Poisson structure for NLS. The other one is new: it chooses purely time-dependent fields as configuration space and yields
a different Poisson structure at each level of the hierarchy.
The corresponding NLEE becomes a {\it space} evolution equation. 
We emphasize the role of the Lagrangian formulation as a unifying framework for deriving both Poisson structures, 
using ideas from covariant field theory. One of our main results is to show that the two matrices of the 
Lax pair satisfy the same form of ultralocal Poisson algebra (up to a sign) 
characterized by an $r$-matrix structure, whereas traditionally only one of them is involved in the classical $r$-matrix method. 
We construct explicit dual hierarchies of Hamiltonians, and Lax representations of the triggered dynamics, 
from the monodromy matrices of either Lax matrix.
An appealing procedure to build a multi-dimensional lattice of Lax pair, through successive uses of the dual Poisson structures, is briefly 
introduced.

\vfill

\noindent {\footnotesize {\tt E-mail: Jean.Avan@u-cergy.fr,\ v.caudrelier@city.ac.uk,\  A.Doikou@hw.ac.uk,\ Anjan.Kundu@saha.ac.in }}\\

\end{titlepage}
\vfill \eject

\tableofcontents

\section{Introduction}

The discovery of the integrability of certain nonlinear evolution equations (NLEE), starting with the pioneering work of Gardner {\it et al.} 
\cite{GGKM} and 
Zakharov-Shabat \cite{ZS}, 
and based on the fundamental notion of a Lax pair \cite{Lax}, has marked a crucial turn in several major domains of mathematical physics 
of the last century. The influence is 
still strongly felt today. Many complementary and interrelated points of view and theories have been developed around this single theme, 
to name but a few:

\begin{itemize}
\item the analytic approach culminating in the inverse scattering approach (see e.g. \cite{AKNS}), 

\item the geometric approach culminating in the Poisson-Lie group formulation of integrable hierarchies via the classical $r$-matrix formalism, 
e.g. \cite{Skly,Drin,STS1,STS2}, 
and the related algebraic approach to integrable hierarchies via Kac-Moody algebras (see e.g. \cite{FNR} and references therein),

\item the algebro-geometric approach dealing with periodic solutions of integrable NLEE, with strong links to Riemann surface theory, e.g. \cite{Nov,Dub,IM}.

\item the pseudo differential operator approach \cite{GD,Dick} applicable in particular to higher-dimensional integrable systems such as KP hierarchy. 
\end{itemize}
The reader will have to excuse our oversimplified list of topics and references 
which cannot make justice to the huge amount of literature accumulated over decades on this theme. Several books can be consulted for a much 
better and fairer description of these areas, e.g. \cite{ft,BBT,BBEIM}.

Despite all these facets, a few key ingredients are always present and provide a unified basis for such treatments: 
a Lax pair formulation and the associated zero curvature representation of the NLEE. The various approaches mentioned above (and the many more hidden within them) are 
motivated by different goals: for instance, the Hamiltonian description of a NLEE is not necessary if one is only interested in viewing it as 
a PDE and finding its solutions. However, it becomes of importance if one insists on proving Liouville integrability of this NLEE viewed as an 
infinite dimensional dynamical system or 
if one has in mind its quantization (one of the key motivations behind the classical $r$-matrix approach). 

We would like to make one simple but striking observation: most of these approaches, and most noticeably the (Liouville) Hamiltonian concept of integrability, always set emphasis on the 
$x$-part of the Lax pair as basic object and view the $t$-part 
as auxiliary. In a sense, this is not at all surprising since the very phrase ``nonlinear evolution equations'' suggests that time is a distinguished 
variable as the parameter
of the evolution flow under investigation.
A departure from this point of view already arises when a given NLEE is viewed as a member of an entire integrable hierarchy describing the commuting flows 
for an infinite number of times $t_n$. The independent variable $x$ is then simply one of the ``times''. However, even with this picture, as soon as 
one singles out a given Lax pair in the hierarchy in view of studying the corresponding {\it evolution} equation, the bias mentioned above is restored and the 
equation is seen as an evolution equation with respect to $t_n$.

In particular, considering NLEE formulated by an AKNS auxiliary problem 
\ba
&&\partial_x \Psi  = {\mathbb U}\,\Psi \label{dif1}\,,
\\ &&
 \partial_t \Psi  = {\mathbb V} \,\Psi\,,
\label{dif2}
\ea
and the related zero curvature condition,
\be
\partial_t {\mathbb U}-\partial_x {\mathbb V} + \Big [ {\mathbb U},{\mathbb V} \Big ] =0,
\ee
and formulating the system in the famous classical $r$-matrix approach to integrable NLEE, 
the fundamental result deals with the Poisson algebra satisfied by the ``fundamental'', $x$-related Lax matrix $\mU$. The $t$-related half of the Lax pair, 
$\mV$, plays a secondary role in that it is determined by the $r$-matrix and the choice of Hamiltonian in the hierarchy deduced from the monodromy of $\mU$.
This point of view is in sharp contrast with the natural symmetric roles played a priori  by $x$ and $t$ in the AKNS formulation.

This symmetry between $x$ and $t$ naturally suggests a notion of duality in the process of choosing $x$ or $t$ as the evolution parameter defining the integrable picture.
Note also that the approach to discretization of conformal integrable quantum theories, e.g. \`a la de Vega-Destri \cite{destri} already
uses as fundamental evolution variables the light cone coordinates $x \pm t$, departing thus from the $x,t$ formulation. 
One may even imagine resorting to some more general (linear) function of $x,t$ as the adequate evolution parameter, depending on the
physical interpretation of the dynamical problem. In other words one may have several relevant choices of classical configuration space, determining the phase space and the Hamiltonian formulation of the dynamical problem. The issue is then whether some other choices than the usual one (here the space slice of fields symbolically 
denoted $\{ \phi(x) \}$) also yield Liouville-integrable Hamiltonian systems. This is the core issue of our investigations.

The need to exploit this simple observation was felt when dealing with initial-boundary value problems as opposed to initial-value problems. 
It was used to its full extent in the development of the so-called unified method of Fokas for initial-boundary value problems \cite{Fokas}. This 
method falls into the analytic category mentioned above.
More recently, still in the realm of initial-boundary value problems, two of the present authors introduced some ideas 
from covariant field theory into the Hamiltonian theory of integrable systems 
to tackle the question of Liouville integrability for a NLEE with an integrable defect \cite{CK}. This approach falls 
into the geometric/Hamiltonian approach mentioned above. It had been motivated by an apparent conceptual gap between earlier studies on 
classical integrability with defects, see e.g. \cite{CZ,VC,Avan-Doikou2} and references therein. 
The main idea was to treat the two independent variables on equal footing when 
developing a classical $r$-matrix approach to integrable defects. This put some of the points made in \cite{AK} on a firm basis. 
As a consequence, on the example of the nonlinear Schr\"odinger (NLS) 
equation, several important observations were made:
\begin{itemize}
\item In addition to the usual Poisson structure $\{~,~\}_S$ used for the Hamiltonian description of NLS as an evolution equation in time, 
one can construct another Poisson structure $\{~,~\}_T$
which describes NLS as an evolution equation in space. Both Hamiltonian descriptions are completely equivalent in that they produce the 
same NLEE but they involve completely different Poisson structures and Hamiltonians. One can speak of a {\it dual picture} whereby 
the same NLEE can be viewed either as a time or space Hamiltonian equation. We see here the explicit example of an alternative choice of the configuration space as 
time-slices of fields $\{ \phi(t) \}$) instead of space-slices. We insist that this scheme must be viewed as completely separated 
from the multitime Hamiltonian procedure where the 
same phase space is endowed with 
a hierarchy of mutually compatible Poisson structures. In this well-known framework the $x$-evolution is triggered by the choice of the 
momentum P as Hamiltonian, and the same Poisson structure. In a nutshell: hierarchies of Poisson bracket structures are mutually compatible 
on the same phase space; dual Poisson structures live on different phase spaces arising from different choices of configuration space, albeit 
for the same original ``dynamical'' system.

\item The second Poisson structure $\{~,~\}_T$ gives rise to an ultralocal Poisson algebra for the $t$-part $\mV$ of the Lax pair, with 
the same classical $r$-matrix (up to a sign) as the ultralocal Poisson algebra satisfied by the $x$-part $\mU$ with respect to the usual 
Poisson bracket $\{~,~\}_S$. We insist that this second Poisson structure is not a compatible Poisson structure in the sense of bi-Hamiltonian theory 
\cite{Magri}. 

\item The Lagrangian formulation becomes the most fundamental description in this context. Indeed, both Poisson structures can be derived from it,
together with the corresponding space and time Hamiltonians, once a choice of general coordinates or configuration space is stipulated. The 
idea of going back to a Lagrangian formulation
of integrable equations of motion, encapsulating a so-called multisymplectic form, goes back at least to the works in \cite{Dick}: at that time 
it was undertaken in the context
of multitime compatible Hamiltonian structures. Here however, as already emphasized, we use the Lagrangian formulation to 
extract two inequivalent choices of configuration space, inducing the construction of different phase spaces and associated Poisson structures 
for our dynamical system.

\end{itemize}

In our opinion, the last observation above is the main conceptual basis for 
an understanding of space-time duality (and possibly more general dualities) in integrable hierarchies. 
Lagrangians naturally contain information on space-time symmetries and do not discriminate between space and time. When one goes over to the 
traditional Hamiltonian picture for a two-dimensional field theory, a distinction between time and space immediately arises once one chooses 
what to call generalized coordinates and their general velocities.

Note that the distinction also occurs, even before Hamiltonian formulation, as soon as
one is lead to specify ``boundary conditions'' for instance distinguishing between $x=cst$ and $t=cst$ boundaries. 
The validity of this scheme and of these observations was also proved in the context of defect theory for the sine-Gordon model by one of 
the authors \cite{C}.
Although the second Poisson structure and its consequences were originally introduced to solve the problem of Liouville integrability 
in the presence of a defect raised in \cite{Avan-Doikou1}, it became clear that the above observations are of a more fundamental nature. 

Let us now specify how we plan to use this scheme.
We shall use both Lagrangian and Hamiltonian formalisms jointly and successively. 
In a first step, the well-known (Hamiltonian-based) construction of a hierarchy of Hamiltonians provides us with a hierarchy of 
simultaneously solvable NLEE. The derivation of a dual Poisson structure then requires to identify the Lagrangian density yielding each NLEE. 
We recover that the Poisson structure describing the $t_n$ evolution is 
the initial Poisson structure, as it should be by first principles of Hamiltonian integrability. The dual Poisson structure describing the 
$x$ evolution at level $n$ must be constructed as a second step by application of the canonical procedure with respect 
to $x$ derivatives of the fields, as dictated by the covariant approach to field theory and the dual choice of phase space and symplectic structure. 
Once it is constructed, the existence of an $r$-matrix structure for the Poisson algebra of the $t_n$ shift operator $\mV^{(n)}$, naturally identified as the Lax operator in this alternative choice, guarantees 
integrability of the second, ``dual'' Hamiltonian evolution. This must be checked by direct calculation.
It is the purpose of this paper to investigate this notion of dual picture, and the procedure just described, in an integrable hierarchy. 
We shall particularize our approach to the nonlinear Schr\"odinger equation as a simple but enlightening example. 
But it is clear that our ideas can be adapted to other AKNS-described hierarchies, with appropriate changes. 

This paper is organised as follows. In Section 2, we give some useful notations and conventions. 
Section 3 is devoted to the introduction of the main concepts of this paper. We describe the dual Hamiltonian formulation
of the NLS hierarchy, the explicit construction being done for level 2 (NLS) and level 3 (complex MKdV). The 
role of the Lagrangian formulation combined with ideas from covariant field theory is emphasized in order 
to derive the two Poisson structures for the dynamical fields. The last part of this section explains how 
the results can be generalized to all the levels in the hierarchy. 
In Section 4, we then build the $r$-matrix formulation associated with this dual picture. 
On top of the usual ultralocal Poisson algebra for the space Lax matrix of NLS, we derive the same Poisson algebra for 
the hierarchy of time Lax matrices (up to a sign). Consequences for conserved quantities and integrability are discussed.
Section 5 is concerned with some conclusions, remarks and definitions based on the results obtained in the previous sections. 
A physical application and an exciting new perspective of future developments are also given. 
Technical issues regarding the construction
of dual hierarchies are given in Appendices A and B.

\section{Notations and conventions}

In the standard approach to the NLS hierarchy in $1+1$ dimensions, the dependent variables (fields) $\psi,\bar\psi$ 
are considered to be functions of an infinite sequence of times $t_n$, $n\ge 0$, with the space variable $x$ being $t_0$ for instance.
We will keep it explicitely as $x$ to make contact with some well-known equations in the NLS hierarchy. 
Each Hamiltonian $H_S^{(n)}$ at level $n$ in the hierarchy describes the evolution of the system with respect 
to a given time $t_n$ in such a way that 
\be
\label{Ham_S}
\psi_{t_n}=\{H_S^{(n)},\psi\}\,,~~\bar\psi_{t_n}=\{H_S^{(n)},\bar\psi\}\,,
\ee
with appropriately chosen Poisson brackets, is equivalent to the zero curvature condition for the Lax pair $(\mU,\mV^{(n)})$ at level $n$
\be
\label{ZCn}
\partial_{t_n}\mU-\partial_x\mV^{(n)}+[\mU,\mV^{(n)}]=0\,.
\ee
In the course of our study, the notion of dual description will emerge whereby, the equations of motion captured by \eqref{ZCn} can 
be equivalently rewritten as the $x$ evolution of fields governed by Hamiltonians that are dual to $H_S^{(n)}$ in a precise sense.
This is why we introduced the subscript $S$ in \eqref{Ham_S} to indicate that the Hamiltonian is a $x$ integral of a local Hamiltonian density in 
the fields and their $x$ derivatives. 
This is of course completely redundant in the traditional approach since this is always assumed to be the case. 
But, in our theory, 
the dual Hamiltonians $H_T^{(n)}$ appear naturally from a Lagrangian picture and 
are now integrals over the formerly identified ``time'' variables $t_n$ of local Hamiltonian densities in the fields and their $t_n$ derivatives. 

The fields and their derivatives with respect to a given subset of the independent variables are denoted as follows
\be
\psi(x,\bt)\,,~~\bar\psi(x,\bt)\,,~~\del_{xt_{j_1\dots j_q}}\psi(x,\bt)=\frac{\del^{q+1} \psi(x,\bt)}
{\del x\del t_{j_1}\dots \del t_{j_q}}\,,~\text{etc.}
\ee
The symbol $\bt$ denotes the sequence of ``times'' (hopefully there is no confusion in using this terminology strongly influenced by the
standard Hamiltonian  description) $t_n$, $n\in\NN$.
When no confusion arises, we omit the arguments altogether. When only some of the independent variables are involved, typically $x$ and 
one $t_n$ or possibly two $t_n$'s, we may display explicitly only those variables that play a role, for instance by writing expressions like 
\be
\psi(x,t_2)\,,~~ \del_{xt_5}\psi=\psi_{xt_5}\,,~~\text{etc.}
\ee

We will use the same rules for all the objects built from the fields and their derivatives (Lagrangians, Hamiltonians, Lax pairs).
Finally, the notion of equal-time Poisson bracket $\{~,~\}^{(n)}_{S}$ and equal-space Poisson brackets $\{~,~\}^{(n)}_{T}$ 
will play an important role in our theory. The subscript $S$ is used to emphasize that the Poisson bracket depends only on the space 
variable ({\it equal-time}) while $T$ is there to remind us that the Poisson depends only on the time variable ({\it equal-space}). 
The superscript encodes the level of the hierarchy that we are looking at.
When writing such Poisson brackets, only the independent variable that are involved will be shown.
For instance, an expression like 
\be
\{\psi(x),\bar\psi(y)\}^{(2)}_{S}=i\delta(x-y)
\ee
involving $\psi(x,\bt)$ and $\bar\psi( y,{\mathbf t^{'}})$ is short for
\be
\{\psi(x,t_2),\bar\psi(y,t_2)\}^{(2)}_{S}=i\delta(x-y)
\ee
and means that we work at equal time (indicated by the subscript $S$) and for $n=2$ (indicated by the supercript $(2)$) \ie $t_2=t^{'}_2$. 
The understanding is that such a Poisson bracket will control the evolution with respect to $t_2$ of a system described 
by a Hamiltonian density depending on $x$. 

In this paper, we consider as boundary conditions the easy-to-handle choice of periodic boundary conditions on a finite interval, 
both in space and time given our purposes. 
Were we to consider field theories on infinite intervals we would introduce the standard fast-decay conditions on the fields. 
More specific boundary conditions on the interval may be considered
for specific purposes, interplaying with bulk integrability properties using suitable formulations drawn from the well-known Cherednik-Sklyanin approaches.

Finally, it is quite profitable to use a notion of scaling dimension to keep track of various quantities in the hierarchy. For instance, we attach 
a scaling dimension $1$ to the fields $\psi$, $\bar\psi$, to the spectral parameter $\lambda$,
 as well as to the derivative with respect to $x$, and a scaling dimension of $n$ 
to the time $t_n$. As we will see in the following, the superscript in $\mV^{(n)}$ captures this notion in that $\mV^{(n)}$ is a polynomial in $\lambda$ 
of the form
\be
\mV^{(n)}(x,\lambda)=\lambda^nV^{(n)}(x)+\lambda^{n-1}V^{(n-1)}(x)+\dots+V^{(0)}(x)
\ee
where $V^{(j)}$ is of scaling dimension $n-j$ and contains only the fields and their $x$-derivatives in combinations of scaling $n-j$. For instance,
in $\mV^{(2)}$, 
$V^{(2)}=\frac{i}{2}\sigma_3$, $V^{(1)}=-\sqrt{\kappa}Q$, with
\ba
Q=\left(\begin{array}{cc}
0 & \bar\psi\\
\psi & 0
\end{array}\right)\,,
\ea
and $V^{(0)}=i\kappa Q^2-i\sigma_3Q_x$. Higher $V^{(j)}$'s can be determined recursively. 
The scaling dimension applies to all other quantities built from the 
fields like the Lagrangian and Hamiltonian densities to be discussed in this paper. Note that, in our notations, a Hamiltonian or Lagrangian 
density with superscript $n$ has in fact scaling dimension $n+2$ as the superscript indicates primarily the level in the hierarchy.

\section{The NLS hierarchy revisited: Lagrangian and dual Hamiltonian pictures}\label{LHP}

\subsection{Lagrangian and dual Hamiltonian picture for NLS}

We review the results first obtained in \cite{CK} in light of the present context and notations. 
A Lax pair for NLS reads
\ba
\mU^{(1)}=-i\frac{\lambda}{2}\sigma_3+\sqrt{\kappa}Q =\begin{pmatrix}
  \frac{\lambda}{2i} & \sqrt{\kappa}\bar \psi \cr
  \sqrt{\kappa}\psi  & -\frac{\lambda}{2i}
\end{pmatrix} \,,
\ea
\ba
\mV^{(2)}&=&i\frac{\lambda^2}{2}\sigma_3-\lambda\sqrt{\kappa}Q(x)+i\kappa Q^2-i\sigma_3Q_x\nonumber\\
&=&\begin{pmatrix}
   -\frac{\lambda^2}{2i} + i \kappa |\psi|^2& -\lambda \sqrt{\kappa} \bar \psi  - i\sqrt{\kappa} \bar \psi_x\cr
  -\lambda \sqrt{\kappa} \psi  + i\sqrt{\kappa}  \psi_x &  \frac{\lambda^2}{2i}- i \kappa |\psi|^2
\end{pmatrix}\,.
\ea
In the rest of this paper, we drop the superscript $(1)$ on $\mU^{(1)}$ as this $\mU$ will be the only one involved here. The zero curvature condition
\be
\partial_{t_2}\mU-\partial_x \mV^{(2)}+[\mU,\mV^{(2)}]=0\,,
\ee
yields the NLS equation,
\be
i\psi_{t_2}+\psi_{xx}=2\kappa|\psi|^2\psi~~\text{(and complex conjugate)}\,.
\ee
A lagrangian density for NLS is 
\ba
\cL^{(2)}=\frac{i}{2}(\bar\psi \psi_{t_2}-\psi\bar\psi_{t_2})-\bar\psi_{x}\psi_{x}-\kappa(\bar\psi \psi)^2\,.
\ea
We now derive the two Poisson brackets $\{~,~\}_S^{(2)}$ (which is the standard one used e.g. in \cite{ft}) and $\{~,~\}_T^{(2)}$, the 
new equal-space Poisson bracket first introduced in \cite{CK}. We will also find the corresponding Hamiltonians $H_S^{(2)}$ and 
$H_T^{(2)}$. For convenience, let us denote $\phi_1=\psi$ and $\phi_2=\bar\psi$ which are treated as independent fields. 
Within this section, since we work at the fixed level $n=2$, it is convenient to 
drop the subscript on $t_2$ and we simply use $t$. Finally, whether we discuss the equal-time picture or the equal-space picture, 
one of the two independent variables will be fixed. Hence, we can afford to use classical mechanics notations instead of field theoretic ones 
(we will then also restrict our attention to densities). By this we mean that instead of writing an expression like
\be
\{\Pi_j(x,t),\phi_k(y,t)\}=\delta_{jk}\delta(x-y)\,,
\ee
we will simply write
\be
\{\Pi_j,\phi_k\}=\delta_{jk}\,.
\ee
Similarly, a quantity like $\del_t \phi_j$ will be written as $\dot{\phi}_j$ and $\del_x \phi_j$ as $\phi_j^{\prime}$.
The context will always be clear so this abuse of notation should not lead to any confusion.
With this in mind, the usual canonical conjugate to $\phi_j$ is defined by
\be	
\Pi_1=\frac{\del \cL^{(2)}}{\del \dot{\phi}_1}=\frac{i}{2}\phi_2\,,~~\Pi_2=\frac{\del \cL^{(2)}}{\del \dot{\phi}_2}=-\frac{i}{2}\phi_1\,.
\ee
These equations can not be used to eliminate $\dot{\phi}_j$ in favour of the momenta $\Pi_j$ when forming the Hamiltonian. On the contrary, we see that they relate 
variables that are supposed to be independent in the Hamiltonian picture \ie $\Pi_j$ and $\phi_k$. So we fall within Dirac's scheme for constrained 
dynamics \cite{Dirac}, 
the primary constraints being
\be
C_1=\Pi_1-\frac{i}{2}\phi_2\,,~~C_2=\Pi_2+\frac{i}{2}\phi_1\,.
\ee
The enlarged Hamiltonian density, taking these constraints into account, is the usual Hamiltonian plus a linear combination of the constraints,
\ba
\cH_S^{*(2)}&=&\cH_S^{(2)}+\lambda_1\,C_1+\lambda_2\,C_2\\
&=&\Pi_1\,\dot{\phi}_1+\Pi_2\,\dot{\phi}_2-\cL^{(2)}+\lambda_1\,C_1+\lambda_2\,C_2\,.
\ea
The initial canonical Poisson brackets are given as usual by
\be
\{\Pi_j,\phi_k\}=\delta_{jk}\,,
\ee
and all the others are zero. The evolution of the constraints under $\cH_S^{*(2)}$ and with respect to these brackets 
must remain zero, which entails the consistency conditions
\be
\{\cH_S^{*(2)},C_j\}=0\,,~~j=1,2\,.
\ee
One computes
\be
\{C_1,C_2\}=-i\,,
\ee
which shows that these primary constraints are second class. Then, the consistency conditions yield
\be
\lambda_1=-i\{\cH_S^{(2)},C_2\}\,,~~\lambda_2=i\{\cH_S^{(2)},C_1\}\,,
\ee
which fixes the value of $\lambda_j$ and shows that there are no more constraints. Dirac's generalized Hamiltonian theory \cite{Dirac} teaches us that 
we can either use $\{~,~\}$ with $\cH_S^{*(2)}$ as determined above to describe the dynamics of the system, or we can use the famous Dirac's bracket 
$\{~,~\}_D$ which allows us to set $C_j=0$ identically, and hence use $\cH_S^{(2)}$ to describe the dynamics. To this end, we need 
the matrix of Poisson brackets of second class constraints
\be
M=\left(\begin{array}{cc}
0 & \{C_1,C_2\}\\
\{C_2,C_1\} & 0
\end{array}\right)=\left(\begin{array}{cc}
0 & -i\\
i & 0
\end{array}\right)\,.
\ee
The Dirac bracket is then defined as
\be
\{f,g\}_D=\{f,g\}-\sum_{j,k=1}^2\{f,C_j\}(M^{-1})_{jk}\{C_k,g\}\,,
\ee
for any functions $f$, $g$ of the dynamical variables. The nonzero canonical brackets with respect to $\{~,~\}_D$ are then modified to 
\be
\label{brackets_NLS1}
\{\phi_1,\phi_2\}_D=i\,,~~\{\Pi_j,\phi_k\}_D=\frac{1}{2}\delta_{jk}\,,~~\{\Pi_1,\Pi_2\}_D=-\frac{i}{4}\,.
\ee
We can see that they are of course consistent with the constraints $C_j=0$ and these are the brackets we should identify with $\{~,~\}_S^{(2)}$. 
With these brackets, we can use the Hamiltonian $\cH_S^{(2)}$ modulo $C_j=0$, which reads
\be
\cH_S^{(2)}=\phi_2^{\prime}\phi_1^{\prime}+\kappa(\phi_1\phi_2)^2\,.
\ee
In terms of the original fields $\psi$, $\bar\psi$, the three brackets \eqref{brackets_NLS1} all consistently reduce to the well-known one \cite{ft}
\be
\{\psi(x),\bar\psi(y)\}_S^{(2)}=i\delta(x-y)\,,
\ee
and $\cH_S^{(2)}$ is the NLS Hamiltonian density $|\psi_x|^2+\kappa|\psi|^4$. The equations of motion arise from
\be
\label{usual_eq_mo}
\psi_t=\{H_S^{(2)},\psi\}_S^{(2)}~~\text{(and complex conjugate)}\,,
\ee
where $H_S^{(2)}$ is the space integral of $\cH_S^{(2)}$.

Following the same principle we now turn to the equal-space Poisson bracket. The idea is to complete the Legendre transform and consider 
another set of canonical conjugate momenta defined with respect to the $x$ derivative of $\phi_j$,
\be	
\cP_1=\frac{\del \cL^{(2)}}{\del \phi^{\prime}_1}=-\phi^{\prime}_2\,,~~\cP_1=\frac{\del \cL^{(2)}}{\del \phi^{\prime}_2}=-\phi^{\prime}_1\,.
\ee
Hence, from the point of view of the $x$ variable, we are in the standard situation where we can eliminate the ``velocities'' $\phi^{\prime}_j$ in 
favour of the momenta $\cP_j$ when forming the Hamiltonian. The analysis is therefore straightforward and we can use the default canonical 
Poisson brackets 
\be
\{\cP_j,\phi_k\}=\delta_{jk}~~\text{(all others zero)}\,,
\ee
as our equal-space bracket $\{~,~\}_T^{(2)}$. The Hamiltonian density is also readily found to be
\be
\cH_T^{(2)}=\cP_1\,\phi_1^{\prime}+\cP_2\,\phi_2^{\prime}-\cL^{(2)}\,.
\ee
In terms of the original fields, this reads
\be
\cH_T^{(2)}=\frac{i}{2}(\psi\bar\psi_{t}-\bar\psi \psi_{t})-\bar\psi_{x}\psi_{x}+\kappa(\bar\psi \psi)^2\,.
\ee
together with\footnote{The difference in signs compared to the results in \cite{CK} comes from the fact that we use the convention $\{p,q\}=1$ 
and $\dot{f}=\{H,f\}$ here, as opposed to the other convention $\{q,p\}=1$ 
and $\dot{f}=\{f,H\}$ used in \cite{CK}. The present convention is in line with that of \cite{ft}.}
\be
\{\psi(t) ,\bar\psi_x(\tau) \}_T^{(2)}=\delta(t-\tau)\,,~~\{\psi(t) ,\bar\psi(\tau) \}_T^{(2)}=0=\{\psi_x(t) ,\bar\psi_x(\tau) \}_T^{(2)}\,.
\ee
The equations of motion are then consistently given by 
\ba
\psi_x=\{H_T^{(2)},\psi\}_T^{(2)}~~\text{(and complex conjugate)}\,,\\
\label{eq_mo_cano}
(\psi_x)_x=\{H_T^{(2)},\psi_x\}_T^{(2)}~~\text{(and complex conjugate)}\,,
\ea
where $H_T^{(2)}$ is the time integral of $\cH_T^{(2)}$. 

As indicated in the Introduction we now have two equally acceptable Poisson structures yielding the same equations of motion, 
albeit with two distinct configuration spaces and unrelated Hamiltonians.

\subsection{Lagrangian and dual Hamiltonian picture for the complex mKdV equation}

At level $(3)$, the zero curvature condition based on the pair $(\mU,\mV^{(3)})$, with
\be
\label{form_V3}
\mV^{(3)}=\begin{pmatrix}
   -{\lambda^3\over 2i} + i \lambda\kappa |\psi|^2+ {\mathrm X} & -\lambda^2 \sqrt{\kappa} \bar \psi  - i\lambda \sqrt{\kappa} \bar \psi_x + \bar  {\mathrm Y} \cr
  -\lambda^2 \sqrt{\kappa} \psi  + i\lambda \sqrt{\kappa} \psi_x +{\mathrm Y}  &  {\lambda^3\over 2i}- i \lambda \kappa |\psi|^2 - {\mathrm X}
\end{pmatrix}
\ee 
and
\be
{\mathrm X} = -\kappa (\bar \psi_x \psi - \psi_x \bar \psi),  ~~ {\mathrm Y} = \sqrt{\kappa}  \psi_{xx} - 2 \kappa^{3\over 2}|\psi|^2 \psi\,,
\ee
yields the complex 
modified KdV equation \cite{CMKDV}.
\be
\label{cmKdV}
\psi_{t_3}-\psi_{xxx}+6\kappa|\psi|^2\psi_{x}=0\,,~~\text{(and complex conjugate)}\,.
\ee
Notice that our $\mV^{(3)}$, as derived from the formulas in Appendix \ref{hierarchy}, 
yields a minus sign in front of the third space derivative, whereas (complex) MKdV is usually presented 
with a plus sign. That is irrelevant as one can simply set $t_3=-t$ and $\kappa\to -\kappa$.

We use the same shorthand notations as before: $t_3=t$ (as we work at the fixed level $n=3$), 
$\dot{\phi}_j$ for $\del_t \phi_j$, $\phi_j^{\prime}$ for $\del_x \phi_j$, etc. A Lagrangian density for this equation is
\ba
\cL^{(3)}&=&\frac{i}{2}(\bar\psi \psi_{t}-\psi\bar\psi_{t})+\frac{i}{2}(\bar\psi_x\,\psi_{xx}-\psi_x\,\bar\psi_{xx})-\frac{3i\kappa}{2}\bar\psi\psi
(\psi\bar\psi_x-\bar\psi\psi_x)\\
&=&\frac{i}{2}(\phi_2 \dot{\phi}_1-\phi_1\dot{\phi}_2)+\frac{i}{2}(\phi_2^{\prime}\,\phi_1^{\prime\prime}-\phi_1^{\prime}\,\phi_2^{\prime\prime})-
\frac{3i\kappa}{2}\phi_2\phi_1
(\phi_1\phi_2^{\prime}-\phi_2\phi_1^{\prime})\,.
\ea

The derivation of the equal-time bracket $\{~,~\}_S^{(3)}$ is exactly the same as in the previous case since the part of $\cL^{(3)}$ involving the 
time derivatives of the fields is identical to that of $\cL^{(2)}$. This is unchanged and this is of course consistent with the fact that we use 
$\mU$ again and only change the $\mV$ part, from $\mV^{(2)}$ to $\mV^{(3)}$. The Lagrangian approach presented is thus consistent with the
traditional Hamiltonian picture where the Poisson bracket is intrinsic to the dynamical system 
(again we are not considering here Magri's multiHamiltonian approach), 
and the change in $\mV$ corresponds to a change in the choice of commuting Hamiltonians. It also means that from now on, we can 
drop the superscript on the equal-time bracket $\{~,~\}_S$. Again, as is well-known from the usual Hamiltonian approach and from what we find 
generically at level $n$ from the Lagrangian picture (see next subsection), the equal-time bracket $\{~,~\}_S$ is the same for the whole hierarchy.

The corresponding Hamiltonian density reads, in terms of the original fields,
\be
\cH_S^{(3)}=-\frac{i}{2}(\bar\psi_x\,\psi_{xx}-\psi_x\,\bar\psi_{xx})+\frac{3i\kappa}{2}\bar\psi\psi
(\psi\bar\psi_x-\bar\psi\psi_x)\,.
\ee

We now turn to the derivation of the equal-space bracket $\{~,~\}_T^{(3)}$ and the corresponding Hamiltonian density. The first step is to define the 
conjugate momenta but we notice that $\cL^{(3)}$ involves not only $\phi_j^{\prime}$ but also $\phi_j^{\prime\prime}$. Therefore we have to deal with 
a higher order Lagrangian theory for which Ostrogradski's construction can be applied. In the following, we use the nice and modern version of 
this construction detailed in \cite{GR}. Another important remark is that, in addition to being of higher order, $\cL^{(3)}$ is also degenerate since the 
higher order derivatives in the fields appear linearly. We have to apply Dirac's scheme as well, along the lines detailed previously for 
obtaining $\{~~\}_S^{(2)}$. Therefore, we adapt the contruction of \cite{GR}, which deals with regular Lagrangians, to our singular case. 
This goes as follows. 
The first step is to reduce the order of the higher order Lagrangian by introducing new fields $\varphi_j$ such that 
\be
\label{constraint1}
\varphi_j=\phi_j^{\prime}\,,~~j=1,2\,.
\ee
The Lagrangian $\cL^{(3)}$ then effectively becomes a standard first order Lagrangian in the independent fields $\phi_j$, $\varphi_j$
\be
\cL^{(3)}=\cL^{(3)}(\phi_j,\dot{\phi}_j,\varphi_j,\varphi_j^{\prime})\,.
\ee
The important relation \eqref{constraint1} ensuring equivalence with the original Lagrangian is now dealt with as a 
constraint by introducing Lagrange multipliers $\mu_j$.
We introduce the following auxiliary Lagrangian 
\be
\cL_{aux}(\phi_j,\dot{\phi}_j,\varphi_j,\varphi_j^{\prime},\mu_j)=\cL^{(3)}(\phi_j,\dot{\phi}_j,\varphi_j,\varphi_j^{\prime})+\sum_{j=1}^2\mu_j(\varphi_j-\phi_j^{\prime})\,.
\ee
For convenience, let us split $\cL^{(3)}$ as
\be
\cL^{(3)}(\phi_j,\dot{\phi}_j,\varphi_j,\varphi_j^{\prime})=\cL_0(\phi_j,\varphi_j,\varphi_j^{\prime})+\cL_1(\phi_j,\dot{\phi}_j)\,,
\ee
with
\ba
\cL_0(\phi_j,\varphi_j,\varphi_j^{\prime})&=&\frac{i}{2}(\varphi_2\,\varphi_1^{\prime}-\varphi_1\,\varphi_2^{\prime})-
\frac{3i\kappa}{2}\phi_2\phi_1
(\phi_1\varphi_2-\phi_2\varphi_1)\,,\\
\cL_1(\phi_j,\dot{\phi}_j)&=&\frac{i}{2}(\phi_2 \dot{\phi}_1-\phi_1\dot{\phi}_2)\,.
\ea
This is motivated by the fact that $\cL_1$ does not play an active role in the construction of $\cH_T^{(3)}$ and the corresponding Poisson brackets, being 
the time part of the Lagrangian. 

We apply the variational principle to the auxiliary Lagrangian for the independent fields $\phi_j$, $\varphi_j$ and $\mu_j$ as usual. Of course, a variation in 
$\mu_j$ enforces the constraints \eqref{constraint1}. A variation in $\phi_j$ yields
\be
\label{eq_motion}
\frac{\del \cL_0}{\del \phi_j}+\mu_j^{\prime}+\frac{\del \cL_1}{\del \phi_j}=\left(\frac{\del \cL_1}{\partial \dot\phi_j}\right)^{\hspace{-0.7cm}\dot \null}~~~~\,.
\ee
A variation in $\varphi_j$ produces an equation fixing $\mu_j$
\be
\label{value_mu}
\mu_j=-\frac{\del \cL_0}{\del \varphi_j}+\left(\frac{\del \cL_0}{\del \varphi^{\prime}_j}\right)^{\prime}\,.
\ee
In particular,
\ba
\mu_1&=&i\varphi_2^{\prime}-\frac{3i\kappa}{2}\phi_2^2\phi_1\,,\\
\mu_2&=&-i\varphi_1^{\prime}+\frac{3i\kappa}{2}\phi_1^2\phi_2\,.
\ea
Note that, upon inserting \eqref{constraint1} and \eqref{value_mu} into \eqref{eq_motion}, we obtain the equations of motion
\ba
&&\dot\phi_1-\phi^{\prime\prime\prime}_1+6\kappa\phi_2^2\,\phi_1=0\,,\\
&&\dot\phi_2-\phi^{\prime\prime\prime}_2+6\kappa\phi_1^2\,\phi_2=0\,,
\ea
that is, \eqref{cmKdV} and its complex conjugate, as required. Note also that \eqref{value_mu} allow us to eliminate $\varphi_j^{\prime}$ in favour 
of the other fields $\mu_j$ and $\phi_j$ when forming the Hamiltonian below.

We can now proceed to the transition to the Hamiltonian picture. We define the conjugate momenta of the fields $\phi_j$, $\varphi_j$ and $\mu_j$,
\be
\cP_j^{0}=\frac{\del \cL_{aux}}{\del \phi_j^{\prime}}\,,~~\cP_j^{1}=\frac{\del \cL_{aux}}{\del \varphi_j^{\prime}}\,,~~\Lambda_j=\frac{\del \cL_{aux}}{\del \mu_j^{\prime}}\,.
\ee
The initial canonical Poisson brackets are therefore
\be
\{\cP_j^{0},\phi_k\}=\delta_{jk}\,,~~\{\cP_j^{1},\varphi_k\}=\delta_{jk}\,,~~\{\Lambda_j,\mu_k\}=\delta_{jk}\,,
\ee
with all others equal to zero. Explicitly, the conjugate momenta read
\be
\label{constraints_Ham1}
\cP_j^{0}=-\mu_j\,,~~\Lambda_j=0\,,
\ee
and
\be
\label{constraints_Ham2}
\cP_1^{1}=\frac{i}{2}\phi_2\,,~~\cP_2^{1}=-\frac{i}{2}\phi_1\,.
\ee
On the Hamiltonian phase space parametrized by $(\phi_j,\cP_j^{0};\varphi_j,\cP_j^{1};\mu_j,\Lambda_j)$, these equations represent primary constraints.
The first set \eqref{constraints_Ham1} is inherent to the construction of \cite{GR}. The second set \eqref{constraints_Ham2} is only present 
in our case because our Lagrangian is singular, in particular, it is linear in the ``velocities'' $\varphi_j^{\prime}$. Note that they are exactly of the 
same nature as the constraints $C_j$ discussed in the NLS case, and they will have exactly the same effect on the Poisson brackets to be derived below. 
We also call them $C_j$ here. Summarizing, we now need to apply Dirac's scheme with the primary constraints
\ba
&&C_1=\cP_1^{1}-\frac{i}{2}\phi_2\,,~~C_2=\cP_2^{1}+\frac{i}{2}\phi_1\,,\\
&&C_3=\cP_1^{0}+\mu_1\,,~~C_4=\cP_2^{0}+\mu_2\,,~~C_5=\Lambda_1\,,~~C_6=\Lambda_2\,.
\ea
We compute the following nonvanishing Poisson brackets among these constraints, which show that they are second class
\be
\{C_1,C_2\}=i\,,~~\{C_3,C_5\}=1=\{C_4,C_6\}\,.
\ee
We now form the Hamiltonian $\cH_{aux}$ in the usual way
\be
\cH_{aux}=\sum_{j=1}^2(\cP_j^{0}\phi_j^{\prime}+\cP_j^{1}\varphi_j^{\prime}+\Lambda_j\mu_j^{\prime})-\cL_{aux}
\ee
as well as the extended Hamiltonian with constraints $\cH^*$
\be
\cH^*=\cH_{aux}+\sum_{j=1}^6 \alpha_jC_j\,.
\ee
The constraints should remain equal to zero under evolution by $\cH^*$ and this yields the consistency conditions
\be
\{\cH^*,C_j\}=0\,,~~j=1,\dots,6\,.
\ee
After straightforward but lengthy calculations, we find that these give
\be
\alpha_j=0\,,~~j=1,2,5,6\,,~~\alpha_j=\left(\frac{\del \cL_1}{\partial \dot\phi_{j-2}}\right)^{\hspace{-0.8cm}\dot \null}~~~~~\,,~~j=3,4\,.
\ee
Thus there are no secondary constraints and the coefficients $\alpha_j$ are fixed. 
The last step is to derive the Dirac brackets for this system so that we can set the constraints to zero identically. The Dirac 
brackets will also be the equal-space brackets $\{~,~\}_T^{(3)}$ that we are looking for.
To this end, we need the matrix of Poisson brackets of the constraints
\be
M=(\{C_j,C_k\})=\left(\begin{array}{cccccc}
0 & i & 0 & 0 & 0 & 0\\
-i & 0 & 0 & 0 & 0 & 0\\
0 & 0 & 0 & 0 & 1 & 0\\
0 & 0 & 0 & 0 & 0 & 1\\
0 & 0 & -1 & 0 & 0 & 0\\
0 & 0 & 0 & -1 & 0 & 0
\end{array}\right)\,.
\ee
As before the Dirac bracket for any two functions $f$, $g$ of the dynamical variables is defined by,
\be
\{f,g\}_D=\{f,g\}-\sum_{j,k=1}^6\{f,C_j\}(M^{-1})_{jk}\{C_k,g\}\,.
\ee
Using these brackets, the constraints can be set to zero identically in the Hamiltonian $\cH^*$ and we can work on the 
reduced phace space consisting of the fields $(\phi_j,\cP_j^{0},\varphi_j)$. We obtain 
\be
\cH_{red}=\cP_1^{0}\varphi_1+\cP_2^{0}\varphi_2+\frac{3i\kappa}{2}\phi_2\phi_1(\phi_1\varphi_2-\phi_2\varphi_1)-\frac{i}{2}(\phi_2\dot\phi_1-\phi_1\dot\phi_2)\,,
\ee
with the Dirac brackets
\ba
\label{Dbrackets1}
&&\{\phi_1,\phi_2\}_D=0\,,~~\{\varphi_1,\varphi_2\}_D=i\,,~~\{\cP^{0}_1,\cP^{0}_2\}_D=0\,,\\
\label{Dbrackets2}
&&\{\phi_j,\varphi_k\}_D=0\,,~~\{\cP^{0}_j,\phi_k\}_D=\delta_{jk}\,,~~\{\cP^{0}_j,\varphi_k\}_D=0\,.
\ea
Restoring all the results in terms of the original fields $\psi$ and $\bar\psi$, we obtain our Hamiltonian density $\cH_T^{(3)}$ after simplification as
\be
\cH_T^{(3)}=i(\bar\psi_x\psi_{xx}-\psi_x\bar\psi_{xx})-\frac{i}{2}(\bar\psi\psi_t-\psi\bar\psi_t)\,.
\ee
The Dirac bracket is the equal-space bracket $\{~,~\}_T^{(3)}$ we want and \eqref{Dbrackets1}-\eqref{Dbrackets2} translate into
\ba
\label{bra1}
&&\{\psi(t),\bar\psi(\tau)\}_T^{(3)}=0\,,\\
&&\{\psi(t),\psi_x(t)\}_T^{(3)}=0=\{\psi(t),\bar\psi_x(\tau)\}_T^{(3)}\,,\\
&&\{\psi_x(t),\psi_{xx}(\tau)\}_T^{(3)}=0=\{\psi_x(t),\bar\psi_{xx}(\tau)\}_T^{(3)}\,,\\
&&\{\psi_x(t),\bar\psi_x(\tau)\}_T^{(3)}=i\delta(t-\tau)\,,\\
&&\{\bar\psi(t),\psi_{xx}(\tau)\}_T^{(3)}=i\delta(t-\tau)\,,~~\{\psi(t),\psi_{xx}(\tau)\}_T^{(3)}=0\,,\\
\label{bra2}
&&\{\psi_{xx}(t),\bar\psi_{xx}(\tau)\}_T^{(3)}=-6i\kappa|\psi|^2(t)\delta(t-\tau)\,.
\ea
One can now verify that 
\be
(\psi_{xx})_x=\{H_T^{(3)},\psi_{xx}\}_T^{(3)}\,,
\ee
and its complex conjugate are indeed equivalent to the complex mKdV equation and its conjugate \eqref{cmKdV}. This concludes our dual Hamiltonian formulation of 
complex mKdV, the equation corresponding to the level $(3)$ in the NLS hierarchy.

The most important outcome in view of our dual classical $r$-matrix approach to be developed in the next section 
is the derivation of the brackets \eqref{bra1}-\eqref{bra2}. Indeed, these allow us to compute the Poisson bracket relations of the 
entries of $\mV^{(3)}$, with the fundamental result that they obey the ultralocal Poisson algebra with rational $r$ matrix, just like $\mU$ does with respect to 
$\{~~\}_S^{(3)}$, but up to a minus sign. 

\subsection{Generalization to higher levels in the hierarchy}\label{gen_higher}

The previous two explicit levels indicate a way in which our scheme will hold for higher levels. 
Let us give here a sketch of this generalization. First of all it is clear that the Lagrangian density at level $n$ is of the form
\be
\cL^{(n)}=\frac{i}{2}(\bar\psi \psi_{t_n}-\psi\bar\psi_{t_n})-\cH_S^{(n)}
\ee
where $\cH_S^{(n)}$ is determined from the generating function $z(L,-L,\lambda)$ via \eqref{z_gen_fct}.
The total number of $x$ derivatives appearing under the integral in $H_S^{(n)}$ is equal to $n$, in view of the recursion relations defining
$W^{(n)}$,
and also consistently with scaling arguments. 
Therefore using integration by parts repeatedly, in view of our assumption of periodic boundary conditions,
we can always bring $\cH_S^{(n)}$ to such a form that the highest order term is proportional to
\be
\begin{cases}
\del_x^p\bar\psi\del_x^p\psi\,,~~n=2p\,,\\
i(\del_x^p\bar\psi\del_x^{p+1}\psi-\del_x^p\psi\del_x^{p+1}\bar\psi)\,,~~n=2p+1\,,
\end{cases}
\ee
NLS is an example of the even case and complex mKdV an example of the odd case, both with $p=1$. For higher values of $p$, the general 
strategy to obtain a full dual Hamiltonian description of the model is clear. The equal-time bracket $\{~,~\}_S$ is the same at all levels 
and is obtained as previously described for NLS or complex mKdV. 
To obtain the equal-space brackets $\{~,~\}_T^{(n)}$ and the corresponding Hamiltonian density $\cH_T^{(n)}$,
we proceed as explained above for NLS and complex mKdV, by using the formalism of \cite{GR}. For even cases, the Lagrangian is regular and the only 
constraints appearing are the ones intrinsic to that higher order formalism. For odd cases, one obtains two more constraints due to the linearity of the Lagrangian 
in the highest $x$ derivative. Irrespective of their origin, all the constraints are second class and no additional constraints appear as a result of the 
consistency conditions. One then simply proceeds as in the complex mKdV case, deriving the Dirac brackets and the corresponding reduced Hamiltonian. These 
are then identified with $\{~,~\}_T^{(n)}$ and $\cH_T^{(n)}$.

\section{Classical $r$-matrix formulation in the dual picture}

\subsection{The original observation: two ultralocal Poisson algebras for NLS}

The main result of the previous section is the derivation from a single Lagrangian picture of two Poisson brackets at each level
 of the NLS hierarchy. Both PBs are indeed represented by an $r$-matrix linear formula for their respective Lax matrix. This guarantees integrability
 of both Hamiltonian dual formulations by construction of monodromy matrices with Poisson-commuting traces. For NLS itself, the result was already obtained and used in \cite{CK} 
 for the specific purpose of 
 studying integrable defects. Let us simply restate the result in the present context.
 
 \begin{proposition}
 Equipped with the canonical Poisson brackets $\{~,~\}_S^{(2)}$ and $\{~,~\}_T^{(2)}$, the Lax matrices $\mU$ and 
 $\mV^{(2)}$ satisfy the following ultralocal Poisson algebras
 \ba
 \label{ultra_PB_U1}
&& \{\mU_1(x,\lambda),\mU_2(y,\mu)\}_S^{(2)}=\delta(x-y)[r_{12}(\lambda-\mu),\mU_1(x_1,\lambda)+\mU_2(y_1,\mu)]\,,\\
\label{ultra_PB_V2}
 &&\{\mV_1^{(2)}(t_2,\lambda),\mV_2^{(2)}(\tau_2,\mu)\}_T^{(2)}=-\delta(t_2-\tau_2)[r_{12}(\lambda-\mu),\mV_1^{(2)}(t_2,\lambda)+\mV_2^{(2)}(\tau_2,\mu)]\,,
 \ea
 where $r$ is the $sl(2)$ rational classical $r$-matrix 
\be
r_{12}(\lambda) = \frac{\kappa}{\lambda}{\mathrm P}_{12}\,,
\ee
${\mathrm P}_{12}$ is the permutation operator acting on $\CC^2\otimes \CC^2$, and we have used the tensor product notation with indices $1$ and $2$.
  
 \end{proposition}
 
It is quite remarkable, and a priori not expected to be a general feature of such duality pictures, that both Hamiltonian structures be described by 
the same $r$ matrix. 
In any case an immediate consequence of the existence of these two Hamiltonian structures is that, in addition to the usual space 
monodromy matrix $\mT_\mU(L,-L,\lambda)$ built from $\mU$ and containing all the information about the system, 
in particular the conserved charges in time, we may now consider the time monodromy matrix $\mT_\mV^{(2)}(T,-T,\lambda)$
built from $\mV^{(2)}$.
In particular, it generates the conserved charges in space. One can talk about 
integrability of the model with respect to these two Poisson structures and hence, about ``dual 
integrable'' hierarchies,  in a sense to be generalized in the Conclusion.
 
\subsection{Hierarchy of dual ultralocal Poisson algebras for higher NLS systems}

At level $(3)$, we may use the Poisson brackets \eqref{bra1}-\eqref{bra2} and the explicit form \eqref{form_V3} for 
$\mV^{(3)}$ to show directly that
 $\mV^{(3)}$ satisfies the ultralocal Poisson algebra \eqref{ultra_PB_V2}
with respect to $\{~,~\}_T^{(3)}$. 

As explained in Section \ref{gen_higher}, one can obtain the higher Poisson brackets 
$\{~,~\}_S^{(n)}$ (which are the same at all levels) and $\{~,~\}_T^{(n)}$ for the fields systematically. It is not hard to convince 
oneself that the ultralocal Poisson algebra structure may hold also for higher $\mV^{(n)}$. 
Hence, we conjecture
\begin{proposition}
 Equipped with the canonical Poisson brackets $\{~,~\}_S^{(n)}$ and $\{~,~\}_T^{(n)}$, the Lax matrices $\mU$ and 
 $\mV^{(n)}$ satisfy the following ultralocal Poisson algebras, for all $n\in\NN$,
 \ba
 \label{ultra_PB_U1n}
 &&\{\mU_1(x,\lambda),\mU_2(y,\mu)\}_S^{(n)}=\delta(x-y)[r_{12}(\lambda-\mu),\mU_1(x,\lambda)+\mU_2(y,\mu)]\,,\\
\label{ultra_PB_Vn}
 &&\{\mV_1^{(n)}(t_n,\lambda),\mV_2^{(n)}(\tau_n,\mu)\}_T^{(n)}=-\delta(t_n-\tau_n)[r_{12}(\lambda-\mu),\mV_1^{(n)}(t_n,\lambda)+\mV_2^{(n)}(\tau_n,\mu)]\,.
 \ea
\end{proposition}

\noindent {\bf Remark}: the Poisson algebra for $\mV^{(n)}$ given in the Proposition is not just a consequence of the Poisson brackets on the dynamical fields involved in 
$\mV^{(n)}$ but is in fact equivalent to them. For $n=3$, this means that the Poisson algebra for $\mV^{(3)}$ is {\it equivalent} 
to the brackets \eqref{bra1}-\eqref{bra2} when we project on the various components. ``Historically'', this is how we discovered 
\eqref{bra1}-\eqref{bra2}: by {\it postulating} 
\eqref{ultra_PB_Vn} for $n=3$ and working out the consequences for the dynamical fieds. This postulate was in turn motivated by the result that was already known
for $n=2$ \cite{CK}. At this level, it was also known that the equal-space brackets for the fields could be {\it derived} from a covariant Lagrangian picture. 
One of our main successes in the present paper is to show that we can complete this picture from the Lagrangian point of view also at the higher levels in the hierarchy, 
as we did explicitly for $n=3$, hence producing \eqref{ultra_PB_Vn} as a nice consequence of the Lagrangian approach. 
However, this ``historical remark'' shows the guiding power of the algebraic approach to Hamiltonian classical integrable systems.

\subsection{A hierarchy of matrices $\mU$ for each $\mV^{(n)}$}\label{Uhierarchy}

Recall that following \cite{ft} (see Appendix \ref{hierarchy}), given the Lax matrix $\mU$ and its ultralocal Poisson algebra \eqref{ultra_PB_U1}, one can 
derive an explicit expression for the generating function of all the matrices $\mV^{(n)}$, $n\in\NN$ of the NLS hierarchy. 
For instance, one finds the first few as:
\ba
&& {\mathbb V}^{(0)}(\lambda) = -{1\over 2i}\begin{pmatrix}
  1& 0 \cr
  0 & -1
\end{pmatrix}\,, \cr
&&  {\mathbb V}^{(1)}(x,\lambda) =\begin{pmatrix}
   -{\lambda\over 2i}& -\sqrt{\kappa} \bar \psi \cr
  -\sqrt{\kappa} \psi & {\lambda \over 2i}
\end{pmatrix}\,, \cr
&& {\mathbb V}^{(2)}(x,\lambda) = \begin{pmatrix}
   -{\lambda^2\over 2i} + i \kappa |\psi|^2& -\lambda \sqrt{\kappa} \bar \psi  - i\sqrt{\kappa} \bar \psi_x\cr
  -\lambda \sqrt{\kappa} \psi  + i\sqrt{\kappa}  \psi_x &  {\lambda^2\over 2i}- i \kappa |\psi|^2
\end{pmatrix}\,,\cr
&& {\mathbb V}^{(3)}(x,\lambda) = \begin{pmatrix}
   -{\lambda^3\over 2i} + i \lambda\kappa |\psi|^2+ {\mathrm X} & -\lambda^2 \sqrt{\kappa} \bar \psi  - i\lambda \sqrt{\kappa} \bar \psi_x + \bar  {\mathrm Y} \cr
  -\lambda^2 \sqrt{\kappa} \psi  + i\lambda \sqrt{\kappa} \psi_x +{\mathrm Y}  &  {\lambda^3\over 2i}- i \lambda \kappa |\psi|^2 - {\mathrm X}
\end{pmatrix}\,, \label{hier1}
\ea
where
\be
{\mathrm X} = -\kappa (\bar \psi_x \psi - \psi_x \bar \psi),  ~~ {\mathrm Y} = \sqrt{\kappa}  \psi_{xx} - 2 \kappa^{3\over 2}|\psi|^2 \psi\,.
\ee

In view of our results, in particular the ultralocal Poisson algebra \eqref{ultra_PB_Vn} satisfied by $\mV^{(n)}$ (proved from $n=1,2,3$, conjectured for higher $n$),
it seems natural to follow the same procedure starting from a fixed $\mV^{(n)}$ and its ultralocal Poisson algebra this time.
Following the construction detailed in Appendix \ref{hierarchy}, we obtain a {\it hierarchy of matrices $\mU^{(m,n)}$, $m\in\NN$} associated to $\mV^{(n)}$.  
As an illustration, suppose we pick $\mV^{(2)}$ corresponding to $\xi=t_2$. We find the first few matrices $\mU^{(m,2)}$ as
\ba
&& {\mathbb U}^{(0,2)}(t_2,\lambda) = {1\over 2i}\begin{pmatrix}
  1& 0 \cr
  0 & -1
\end{pmatrix}\,, \cr
&&  {\mathbb U}^{(1,2)}(t_2; \lambda) =\begin{pmatrix}
   {\lambda\over 2i}& +\sqrt{\kappa} \bar \psi \cr
  \sqrt{\kappa} \psi & -{\lambda \over 2i}
\end{pmatrix}\,, \cr
&& {\mathbb U}^{(2,2)}(t_2;\lambda) = \begin{pmatrix}
   {\lambda^2\over 2i} - i \kappa |\psi|^2& +\lambda \sqrt{\kappa} \bar \psi  + i\sqrt{\kappa} \bar \psi_x\cr
  \lambda \sqrt{\kappa} \psi  - i\sqrt{\kappa}  \psi_x &  -{\lambda^2\over 2i}+ i \kappa |\psi|^2
\end{pmatrix}\,,\cr
&& {\mathbb U}^{(3,2)}(t_2;\lambda) = \begin{pmatrix}
   {\lambda^3\over 2i} - i \lambda\kappa |\psi|^2- \Phi & \lambda^2 \sqrt{\kappa} \bar \psi  +i\lambda \sqrt{\kappa} \bar \psi_x - \bar  \Omega \cr
  \lambda^2 \sqrt{\kappa} \psi  - i\lambda \sqrt{\kappa} \psi_x -\Omega  &  -{\lambda^3\over 2i}+i \lambda \kappa |\psi|^2 + \Phi
\end{pmatrix}\,,
\ea
where we define
\be
 \Phi= -\kappa (\bar \psi_x \psi - \psi_x \bar \psi),  ~~ \Omega = -i\sqrt{\kappa}  \psi_{t_2} - 2 \kappa^{3\over 2}|\psi|^2 \psi
\ee
Note the appearance of a derivative with respect to $t_2$ in ${\mathbb U}^{(3,2)}$ in addition to the $x$-derivatives. 

Once this hierarchy is constructed, we can in principle repeat the process starting from a given $\mU^{(m,n)}$ and obtain a hierarchy 
of associated Lax matrices $\mV^{(p,m,n)}$, $p\in\NN$, and so on.

This appealing idea suggests that one can construct a collection of Lax pairs $(\mU^{(I)},\mV^{(J)})$, where $I,J$ are 
multiindices, which could be organised in a multidimensional lattice, hence producing multidimensional integrable PDEs. 
One is naturally led to distinguish between ``vertical'' and ``horizontal'' indices 
depending whether the index labels a hierarchy of elements built from a ``equal-space'' or ``equal-time'' Poisson bracket. Quite obviously the even (resp. odd)
indices of $\mV$ (resp $\mU$ ) are horizontal indices in this sense.
The detailed investigation of such a possibility and of the corresponding lattice of Lax pairs is left for future work. 
A very special example of this idea appeared differently in the form of a triple of Lax matrices in \cite{KM} and was at the basis of a certain 
$2+1$-dimensional version of NLS. In general, it is far from obvious 
that the process just described ``closes'' in the sense that the lattice is finite dimensional and that all the equations obtained by taking the 
zero curvature associated to each pair in the lattice are mutually consistent. The last issue is already a nontrivial result in the well-known case of a single 
hierarchy (a one-dimensional lattice from our point of view) that can be found e.g. in \cite{FNR}

\section{Conclusions and outlook}

Our main results can be summarized as follows:

$\bullet$ {\bf Lagrangian and dual Hamiltonian formulation:} For each Lax pair $(\mU,\mV^{(n)})$ at level $n$ in the NLS hierarchy, we construct two Poisson 
brackets $\{~,~\}^{(n)}_S$ and $\{~,~\}^{(n)}_T$ 
for the canonical fields and two Hamiltonians $H_S^{(n)}$ and $H_T^{(n)}$ (recall again that these are not Magri's multi-Hamiltonian structures).
They are derived from a Lagrangian which reproduces the equations of motion 
of the zero curvature condition for $(\mU,\mV^{(n)})$ through its Euler-Lagrange equations. The equal-time bracket $\{~,~\}^{(n)}_S$ 
found in this procedure is the same at each level $n$, which is consistent with what is known from the traditional Hamiltonian approach to an integrable 
hierarchy .

$\bullet$ {\bf Dual classical $r$-matrix formulation:} For each Lax pair $(\mU,\mV^{(n)})$ at level $n$ in the NLS hierarchy, the Poisson bracket 
$\{~,~\}^{(n)}_S$ (resp. $\{~,~\}^{(n)}_T$) for the canonical fields induces an ultralocal Poisson algebra for $\mU$ (resp. $\mV^{(n)}$). 
A consequence of this is that the space monodromy matrix $\mT_\mU$ generates integrals of motion (in time) that are in involution with respect to $\{~,~\}^{(n)}_S$
(the well-known result) and the time monodromy matrix $\mT_{\mV^{(n)}}$ generates integrals of motion (in space) 
that are in involution with respect to $\{~,~\}^{(n)}_T$ (the new, dual result). The Hamiltonians $H_S^{(n)}$ and $H_T^{(n)}$ are 
contained in these families of integrals of motion.

We are now in a position to formulate some general remarks and more precise definitions of our procedures.

\subsection{The notion of duality in an integrable hierarchy}

We have used the terminology {\it dual} or {\it duality} rather loosely so far, counting on the fact that it is self-explanatory in the various 
examples that we have used to illustrate it. However, we would now like to propose an attempt at a more precise definition to be used
in future developments.

We recall that the general feature of a Lax representation of AKNS type in Hamiltonian formalism is as follows.
Consider a Lax matrix $\mU(\phi(x), \lambda)$, depending on one-dimensional fields $\phi(x)$ living on the phase space
and a spectral parameter $\lambda$ and taken to describe the infinitesimal $x$-translation 
for an auxiliary wavefunction; and consider a Poisson structure for these fields $\phi$ inducing an $r$-matrix structure for the Lax matrix.

The monodromy $\mT_\mU$ of this Lax matrix then naturally generates Poisson commuting Hamiltonians as integrals over $x$ of (local) densities, 
and the data of $\mU(\phi(x), \lambda)$ and $r$ allow one to construct an $\mV$ matrix 
 depending on $\lambda$, $\phi$ and possibly the $x$-derivatives of $\phi$, for every such Hamiltonian, 
 describing a time translation for the auxiliary wavefunction, and such that the compatibility of the $t$- and $x$- evolutions (flatness of connection) 
 implies the Hamiltonian equations of motion (time evolution) for the initial fields 
 $\phi(x)$ now viewed as $\phi(x,t)$. The Poisson structure is here by construction an equal-time Poisson bracket.

In view of the explicit results obtained so far on our NLS example, we shall define the notion of {\it dual formulation} as follows.
Suppose one can exhibit another, {\it equal-space} Poisson structure for the fields, now chosen to be $\phi(t)$, 
parametrizing another choice of configuration space such that:

- The corresponding Poisson structure of the matrix $\mV$ now naturally identified as the relevant Lax matrix, 
has a linear $r$-matrix description with matrix $r'$, and, 

- the $\lambda$ expansion of the monodromy matrix $\mT_\mV$ of the $t$-translation described by $\mV(\phi(t),\lambda)$
contains in particular a Hamiltonian integral over $t$, such that
its associated translation operator built from the $r$-matrix structure $r'$ as in Appendix \ref{hierarchy} coincides with $\mU$.

Then the dynamical equation for $\phi$, contained in the zero-curvature condition for $\mU, \mV$, can be either represented as a Hamiltonian 
integrable evolution for $\phi(x)$ along $t$ or for $\phi(t)$ along $x$. 

This definition is here explicitely formulated for AKNS systems describing mutually compatible $x$ and $t$ evolutions. It must be emphasized that one may
easily extend this notion to any pair of AKNS-type gauge connections over a two-dimensional space-time, originally yielding by zero-curvature 
condition 2-dimensional nonlinear evolution equations for some fields.
As above, in the general case these connections will be respectively identified with the infinitesimal shifts of an auxiliary wavefunction, this time 
along two independent space-time coordinates, linear combinations or even
algebraic combinations of $x$ and $t$ : the example of light-cone coordinates mentioned above is one obvious such extension.

Whenever such a scheme is realised, we shall speak of {\it dual (canonical) formulations} or {\it dual (canonical) representations} 
of the integrable PDE. As is seen on the definition we impose 
in particular that in both cases the ``generalized time-shift'' element of the Lax connection  will be identified as a partial trace describing an
$r$-matrix co-adjoint action on the differential of the Hamiltonian derived from interval monodromies of the ``generalized space-shift'' element, 
such as derived in \cite{STS1}. This implies a very non trivial
property on the ``initial'' Lax matrix (be it picked as $\mU$ or $\mV$), i.e. that it must be recovered from $r$-matrix coadjoint action
on the spectral-parameter expansion of the monodromy matrix of its Lax partner which is itself obtained from $r$-matrix coadjoint action
on the spectral-parameter expansion of the first matrix; a very nontrivial consistency of the set (Lax matrix / $r$-matrix) is thus suggested.

A general caveat is in order here: The identification of the second partner of a Lax formulation with the coadjoint formula \`a la Semenov-Tjan-Shanskii
requires that the particular Hamiltonian triggering the dynamics of the considered integrable system belong to the ring generated by the trace
of the monodromy matrix of the first partner, or the spectral-parameter expansion coefficients thereof. In other words it is strictly speaking
a sufficient condition statement. However there exists situations where the Hamiltonian may not necessarily be constructed from
the monodromy matrix of the chosen Lax matrix: such situations certainly occur in the context of ``degenerate integrable systems'' \cite{resh} and may
also occur in the 
context of field-theoretical models where the existence of an infinite number of degrees of freedom requires extra completeness arguments. We shall assume that 
we are not in such a situation.

In the case where the $r$-matrices coincide, $r=r'$, we shall call the system a self-dual Hamiltonian system. We do not have examples at this time
of dual integrable systems with different $r$ matrices for the two dual integrability structures but we cannot exclude such situations,
hence this particularization. Note that the NLS equation can then be called anti-self-dual since we showed that we have $r'=-r$ in this paper.

The definition can be easily extended to a whole hierarchy. Once again, given a Lax matrix $\mU$ and its Poisson algebra characterised by 
$r$, one can build a whole integrable hierarchy of Lax pairs $(\mU,\mV^{(n)})$, $n\in\NN$, in which $\mV=\mV^{(1)}$ say. Then, by 
the notion of duality just explained, we have the data of $\mV$ and its associated Poisson algebra characterised by $r'$. Therefore, we can build 
a hierarchy of Lax matrices $\mU^{(n)}$, $n\in\NN$, with $\mU^{(1)}=\mU$. The hierarchies $(\mU^{(1)},\mV^{(n)})$, $n\in\NN$ and 
$(\mU^{(n)},\mV^{(1)})$, $n\in\NN$ are called {\it dual hierarchies}.

Note that the terminology of ``duality'' has already appeared in the domain of integrable systems, albeit in a different context, 
i.e. the famous position/momentum duality before Hamiltonian reduction,
relating Calogero-Moser and Ruijsenaars-Schneider models under the generic name of Ruijsenaars duality \cite{Rui,Feh}, with identical Poisson structure. 
Here by contrast 
the ``duality'' exchanges the role of variables $x$ and $t$ in the definition of integrability itself, and the Poisson structures are distinct. 
In a sense and to borrow a phrase from string theory, RS duality is target space duality, whereas our notion is more related to world sheet duality, or change of the conformal structure.

\subsection{Physical application: a Hamiltonian description of NLS in Optics}

The NLS equation that served as the main basis to develop our formalism turns out to be also an excellent physical example of 
the interchange between space and time in certain integrable systems with a universal character. By this we mean that 
NLS arises as an approximate equation in wave dynamics under fairly general assumptions but with different interpretations of the 
independent variables depending on the context. A nice discussion of the physical significance and universal character of NLS can be found in 
\cite{SS}. When derived in the fluid mechanics context for instance, the usual form is relevant
\be
\label{standard}
i\phi_t+\phi_{xx}=2\kappa|\phi|^2\phi\,,
\ee
where $t$ has the meaning of time and $x$ the meaning of space. That is by far the most popular setting and the standard 
Hamiltonian description is always given for this form.
But, when NLS is derived in nonlinear optics, the physical meaning attached to $x$ and $t$ is swapped. For instance, 
in Chapter 5 of \cite{Ag}, one finds NLS in the form
\be
iA_z=\frac{\beta_2}{2}\,A_{\tau\tau}-\gamma|A|^2A\,,
\ee
where $A(z,\tau)$ is the amplitude of the pulse envelope and $\beta_2, \gamma$ are parameters characterizing the physical properties of the optical fiber.
Here $z$ is the spatial coordinate along the fiber and $\tau$ is attached to a frame of reference moving with the pulse at group velocity $v_g$, 
$\tau=t-\frac{z}{v_g}$. Of course this can always be rescaled to the standard form \eqref{standard} but the point is that $x$ is now a time 
and $t$ a space variable. To implement the Hamiltonian approach of this form of NLS seen as an evolution equation {\it in time}, one has
in fact to use our formalism to produce $\phi_{xx}=(\phi_{x})_x$ as the Poisson bracket of some Hamiltonian with $\phi_x$. This is nothing but 
\eqref{eq_mo_cano}.

This example is also illuminating to motivate the need for the notion of Hamiltonian space evolution that we have advocated for in this paper. 
Indeed, suppose that instead of using our new bracket as just described, 
one insists on using the standard Hamiltonian formulation of \eqref{standard} \ie \eqref{usual_eq_mo}, but now in the physical context 
of optical fibers. Then one is in fact describing the {\it space} evolution of the wave.

This discussion shows that even in the nonrelativistic setting, it is sometimes useful not to attach an intrinsic meaning 
to what coordinate describes space and what coordinate describes time. We stress once again that, be it mathematically or physically, 
treating variables on an equal footing in a given integrable hierarchy provides new insight. 

\subsection{Multi-index Lax pairs}

We have suggested in the previous section that one could imagine successive repeats of the dualization procedure, starting from the two initial hierarchies of
$\mU$ and $\mV$, and successively building higher levels of ``horizontal'' or $\mU$-type and ``vertical'' or $\mV$-type hierarchies by starting from a lower level
hierarchy element, assuming it is endowed with the double set of Poisson structures. An immediate question arises whether there exists connections between
$\mU$ and $\mV$ objects respectively labeled by their ordered set of alternatively ``vertical'' and ``horizontal'' indices: $(\mU^{(I)},\mV^{(J)})$ when some relation
exists between the respective sets $I$ and $J$ (a more precise statement of the issue of ``finite-dimensionality of the lattice'' alluded to in the previous section).
An obvious second question is whether the dual Poisson structures a priori deduced from the Lagrangian formulation of higher-level hierarchy equations are still
represented by $r$-matrix formulations; whether these $r$-matrices will then remain the same (as is the case
for the first two levels of hierarchy in NLS); or whether new $r$-matrices will arise; or (worst-case) dual integrability breaks altogether at higher levels
and no dual $r$-matrix formulation is available.

\section*{Ackowledgements}

Preliminary steps for this work were undertaken while J.A. visited City University London and while V.C. visited Heriot-Watt University. 
The Indian National Science Academy is warmly ackowledged for awarding the Dr Ramalingaswami INSA Chair 2015 to V.C. and supporting his visit 
to the Bose Centre and Saha Institute in Kolkata, during which the main part of this work was completed.

\section*{Appendix} 
\appendix

\section{Hierarchy of Lax matrices from the $r$-matrix}\label{hierarchy}

Given the symmetric roles played by the matrices $\mU$ and $\mV^{(n)}$ and in view of the discussion in Section \ref{Uhierarchy}, we 
present in this appendix the generalization of the construction of \cite{ft} [Chap III, $\S$3]. It yields a formula for a generating function 
of the hierarchy of Lax matrices $\mV^{(n)}$ given $\mU$ and the classical $r$-matrix encoding the Poisson algebra of the latter. 
For our purposes, we would like to be able to base the construction on any of the $\mV^{(n)}$, and the corresponding $r$-matrix encoding 
its Poisson algebra, in order to obtain a hierarchy of Lax matrices $\mU^{(m)}$ associated to this $\mV^{(n)}$. Ultimately, we 
can also imagine to pick one of the $\mU^{(m)}$ thus constructed, compute its Poisson algebra and repeat the construction once again. 
The study of the iteration of this process and of the lattice of Lax pairs that one can obtain this way will be the object of a future investigation.

We need neutral notations to allow for this process. So, let us 
assume that we have constructed appropriate Poisson brackets $\{~,~\}^{(N)}_{A}$\footnote{$A$ 
can stand for either $S$ or $T$ here depending on 
whether we use $\xi$ as space and $\eta$ as time or vice versa. $N$ is a fixed level that we take as a starting point.},
 such that equations of motion arising from the zero curvature for the
Lax pair $(\mX,\mY)$
\ba
\label{ZC_level_N}
\del_{\eta_N} \mX(\xi,\eta_N,\lambda)-\del_{\xi}\mY(\xi,\eta_N,\lambda)+[\mX,\mY](\xi,\eta_N,\lambda)=0\,,
\ea
 read
\ba
\del_{\eta_N}\mX(\xi,\eta_N,\lambda)=\{H^{(N)},\mX(\xi,\eta_N,\lambda)\}^{(N)}_{A}\,.
\ea
Here, $\mX$, $\mY$ matrix polynomials in $\lambda$ of degree $N$ and $M$ respectively, and 
\ba
Tr\,\mX=0~~,~~\bar\mX(\xi,\eta,\bar\lambda)=\sigma\,\mX(\xi,\eta,\lambda)\,\sigma\,,
\ea
where $\sigma=\sigma_1$ if $\kappa>0$ and $\sigma=\sigma_2$ if $\kappa<0$.
We then assume that these equations of motion at the single level $N$ are embedded into the following hierarchy of equations
\ba
\del_{\eta_n}\mX(\xi,\eta_n,\lambda)=\{H^{(n)},\mX(\xi,\eta_n,\lambda)\}^{(N)}_{A}\,,~~n\in\NN\,,
\ea
where the Hamiltonians $H^{(n)}$ are constructed from $\mX$ as in Appendix \ref{Ricc}.
In addition, assume that 
\ba
\label{r_bracket_gamma}
\{\mX_1(\xi,\eta,\lambda),\mX_2(\zeta,\eta,\mu)\}^{(N)}_{A}=\gamma\delta(\xi-\zeta)[r_{12}(\lambda-\mu),\mX_1(\xi,\eta,\lambda)+\mX_2(\zeta,\eta,\mu)]\,,
\ea
where $\gamma=\pm 1$ and $r$ is the standard one-pole rational classical $r$-matrix . This last assumption is motivated by one of the main results of this 
paper whereby $\mU$ and $\mV^{(n)}$ satisfy the same Poisson algebra up to a sign, hence the introduction of $\gamma$.

At this stage, we can repeat the arguments of \cite{ft} [Chap III, $\S$3] and simply incorporate $\gamma$.
We obtain the equations of motion in zero curvature form for each level $n$ as
\ba
\del_{\eta_n} \mX(\xi,\eta_n,\lambda)-\del_{\xi}\mY^{(n)}(\xi,\eta_n,\lambda)+[\mX,\mY^{(n)}](\xi,\eta_n,\lambda)=0\,.
\ea
where the hierarchy of matrices $\mY^{(n)}$ is obtained from the expansion
\ba
\mY(\xi,\eta,\lambda;\mu)=\kappa\sum_{n=1}^\infty\frac{\mY^{(n-1)}(\xi,\eta,\lambda)}{\mu^n}\,,
\ea
where
\ba
\label{expression_generating_hierarchy}
\mY(\xi,\eta,\lambda;\mu)=\frac{\gamma\kappa}{2i(\lambda-\mu)}(\1+W(\xi,\eta,\mu))\,\sigma_3\,(\1+W(\xi,\eta,\mu))^{-1}\,.
\ea
We have added the explicit $\eta$ dependence of $W$ (defined in \eqref{form_T}) for consistency. In particular, we obtain that $\mY^N$ coincides 
with the original $\mY$ in \eqref{ZC_level_N}, as it should. 

Of course, if we apply this construction to $\mX=\mU$ with $\xi=x$ and $\gamma=1$, we recover $\mY^{(n)}=\mV^{(n)}$ with $\eta_n=t_n$. The idea 
is that we want to be able to apply the construction to $\mX=\mV^{(n)}$ for some fixed $n$ and $\xi=t_n$, to yield a hierachy of $\mU^{(m)}$'s, etc.

{\it Remark:} An important consequence of formula \eqref{expression_generating_hierarchy} 
is that the two properties of $\mX$, \ie tracelessness and symmetry, transfer to all the 
$\mY^{(n)}(\xi,\eta,\lambda)$'s. This is crucial if we want to be able to repeat the construction with one of the $\mY^{(n)}$'s as $\mX$ since 
these properties enter the argument to obtain formula \eqref{expression_generating_hierarchy}. 

We now investigate a consequence of formula \eqref{expression_generating_hierarchy} that gives a useful recursive formula 
for $\mY^{(n)}$ in terms of $\mY^{(n-1)}$ and the coefficients $W^{(j)}$, $j=1,\dots,n$ discussed in Appendix \ref{Ricc}. 
To our knowledge, this recursion formula was not known (at least not given) before.
It reads,
\ba
&&\mY^{(0)}(\xi,\eta,\lambda)=\frac{i\gamma\kappa}{2}\sigma_3\,,\\
\label{recurrence_Lax_matrix}
&&\mY^{(n)}(\xi,\eta,\lambda)=\lambda\mY^{(n-1)}(\xi,\eta,\lambda)+i\gamma\sigma_3\sum_{j=1}^n(-1)^j\sum_{m_1+\dots+m_j=n}W^{(m_1)}\dots W^{(m_j)}\,.
\ea
In the last sum, for a given $j$, the sum runs over all partition of $n$ into $j$ integers $m_1,\dots,m_j$. To prove it, 
simply multiply \eqref{expression_generating_hierarchy} by $\lambda-\mu$ and rewrite it as
\ba
\left[\sum_{n=1}^\infty\frac{1}{\mu^n}\left(\lambda\mY^{(n-1)}-\mY^{(n)}\right)-\mY^{(0)}-i\frac{\gamma}{2}\sigma_3  \right]=-i\gamma\sigma_3 (\1+W)^{-1}\,.
\ea 
The result follow by identifying the term of order $\mu^{-n}$ on each side.

\section{Riccati equation and conserved quantities}\label{Ricc}
We keep the notations introduced in the previous Appendix that allow us to discuss the construction of a generating function 
of conserved quantities in time or space, depending on the starting point used for $\mX$ and $\xi$ as explained above. Again, we 
extend the arguments presented in \cite{ft} [Chap I, $\S$4] to our case. The ensuing results are obtained by direct calculation.
Consider the auxiliary problem
\ba
\label{Lax1}
&&{\partial_\xi}\Psi(\xi,\eta, \lambda) = {\mathbb X}(\xi,\eta, \lambda) \Psi(\xi,\eta, \lambda)\\ 
\label{Lax2}&&
{\partial_\eta }\Psi(\xi,\eta, \lambda) = {\mathbb Y}(\xi,\eta,\lambda) \Psi(\xi,\eta, \lambda)
\ea
with $\mX$, $\mY$ matrix polynomials in $\lambda$ of degree $N$ and $M$ respectively, and 
\ba
Tr\,\mX=0~~,~~\bar\mX(\xi,\eta,\bar\lambda)=\sigma\,\mX(\xi,\eta,\lambda)\,\sigma\,,
\ea
where $\sigma=\sigma_1$ if $\kappa>0$ and $\sigma=\sigma_2$ if $\kappa<0$. 
Consider periodic boundary conditions in $\xi\in[-\Lambda,\Lambda]$ \footnote{Note that in general $\xi$ will be either a space $x$ or a time $t_n$ 
so that we work with periodic conditions both in space and time variables. 
The well-known finite gap solutions fulfil this requirement.}. Let $\mT(\xi,\zeta,\lambda)$ be the fundamental solution of \eqref{Lax1} normalised to $\mT(\xi,\xi,\lambda)=\1$. Introduce the 
large $\lambda$ expansion of $\mT$ as follows
\ba
\label{form_T}
\mT(\xi,\zeta,\lambda)=(\1+W(\xi,\lambda))e^{Z(\xi,\zeta,\lambda)}(\1+W(\zeta,\lambda))^{-1}\,,
\ea
where $W$ is an off-diagonal matrix, $Z$ a diagonal matrix, and they have the following expansion
\ba
W(\xi,\lambda)=\sum_{n=1}^\infty\frac{W^{(n)}(\xi)}{\lambda^n}~~,~~Z(\xi,\zeta,\lambda)=\sum_{n=0}^N Z^{(-n)}(\xi,\zeta)\,\lambda^n+\sum_{n=1}^\infty\frac{Z^{(n)}(\xi,\zeta)}{\lambda^n}\,.
\ea 
A generating function of the real-valued local conserved quantities in $\eta$ is now given by
\be
\label{generating}
\frac{1}{2i\kappa}Tr[\sigma_3 Z(\Lambda,-\Lambda,\lambda)]
\ee
with 
\ba
Z(\xi,\zeta,\lambda)= \int_{\zeta}^\xi\left(\mX_d+\mX_o\,W\right)(z,\lambda)\,dz\,,
\ea
where $\mX_d$ (resp. $\mX_o$) is the diagonal (resp. off-diagonal) part of $\mX$. $W$ is determined via the matrix Riccati equation 
\be
W_\xi=\mX_d\,W-W\,\mX_d+\mX_o-W\,\mX_o\,W\,.
\ee
The Hamiltonians are obtained by inserting the following expressions for $Z^{(n)}(\xi,\zeta)$
\ba
&&Z^{(-n)}(\xi,\zeta)=\int_\zeta^\xi\left(\mX_d^{(n)}(z)+\sum_{p=n+1}^N\mX_o^{(p)}(z)\,W^{(p-n)}(z)\right)dz\,,~~n=0,\dots,N\,\\
&&Z^{(n)}(\xi,\zeta)=\sum_{p=0}^N\int_\zeta^\xi\left(\mX_o^{(p)}(z)\,W^{(p+n)}(z)\right)dz\,,~~n\ge 1\,.
\ea
and finding the coefficients $W^{(n)}$ recursively thanks to, for $n=1,\dots,N$, 
\ba
\label{recurrence_W_lower}
[\mX_d^{(N)},W^{(n)}]=-\mX_o^{(N-n)}-\sum_{q=1}^{n-1}[\mX_d^{(N-q)},W^{(n-q)}]-\sum_{m=1}^n\sum_{k=1}^{m-1}W^{(k)}\,\mX_o^{(N+m-n)}\,W^{(m-k)}\,.
\ea
and, for $n\ge 1$,
 \ba
 \label{recurrence_W_upper}
[\mX_d^{(N)},W^{(N+n)}]=W^{(n)}_\xi-\sum_{k=1}^{N}[\mX_d^{(N-k)},W^{(N+n-k)}]-\sum_{p=0}^N\sum_{k=1}^{n+p-1}W^{(k)}\,\mX_o^{(p)}\,W^{(n+p-k)}\,.
\ea
Note that $\mX_d^{(N)}$ is always proportional to $\sigma_3$. The previous Appendix shows that this is a stable property within a hierarchy. 
Hence these relations define $W^{(n)}$ fully, as it is off-diagonal.
As is well-known, the fact that $\mX$ is traceless and the normalisation 
of $\mT$ imply that 
\be
\det \mT(\xi,\zeta,\lambda)=1\,.
\ee
In addition, using the periodic ``boundary'' conditions and the decomposition \eqref{form_T} for large $\lambda$, one obtains
\ba
\label{conservation}
\del_\eta Tr[e^{Z(\Lambda,-\Lambda,\lambda)}]=0\,.
\ea
From the recursive definitions of the coefficients $W^{(n)}$ and $Z^{(n)}$ in terms of $\mX$, we can see that the symmetry condition on the latter is transferred to 
all these coefficients, and hence also to the functions $W(\xi,\lambda)$ and $Z(\xi,\zeta,\lambda)$. Hence we can write 
\ba
W^{(n)}(\xi)=i\sqrt{\kappa}\left(\begin{array}{cc}
0 & -\bar w^{(n)}(\xi)\\
w^{(n)}(\xi) &0
\end{array}\right)~~,~~Z^{(n)}(\xi,\zeta)=i\kappa z^{(n)}(\xi,\zeta)\sigma_3\,,~~z^{(n)}\in\RR\,.
\ea
In particular, this yields
\ba
z(\Lambda,-\Lambda,\lambda)=\frac{1}{2i\kappa}Tr[\sigma_3 Z(\Lambda,-\Lambda,\lambda)]\,.
\ea
In addition, $Tr[e^{Z(\Lambda,-\Lambda,\lambda)}]=2i\kappa \cos(z(\Lambda,-\Lambda,\lambda))$ so \eqref{conservation} implies that 
\be
\del_\eta z(\Lambda,-\Lambda,\lambda)=0\,.
\ee
Once again, the recursion relations show that this function generates {\it local} conserved quantities as opposed to $Tr[e^{Z(\Lambda,-\Lambda,\lambda)}]$.
We have also shown that all the $z^{(n)}$'s are real valued. The corresponding hierarchy of local Hamiltonians is now simply defined by 
\be
\label{z_gen_fct}
H^{(n)}=z^{(n+1)}(\Lambda,-\Lambda,\lambda)=\int_{-\Lambda}^\Lambda\cH^{(n)}(\xi)\,d\xi\,.
\ee
They generate the evolution with respect to the variable $\eta_n$ (see Appendix \ref{hierarchy}).

\end{document}